\documentclass[9pt]{article} 
\usepackage[T1]{fontenc}
\usepackage{graphicx}
\usepackage{bookman}
\usepackage{latexsym}
\usepackage{amssymb}
\usepackage{amsbsy}
\usepackage{cite}
\begin{document}
\title{Chen-Wu type model and the total energy of the universe}
 \author{M.de Campos$^{(1)}$}
\maketitle
\footnote{
{ \small \it UFRR
 Campus do Paricar\~ana.  Boa Vista. RR. Brasil. 
 campos@dfis.ufrr.br }} 
\begin{abstract}
The inclusion of a $\Lambda $ term, today, in the relativistic field equations is consequence of the evolution of an observational cosmology.
In this work we argue the "ansatz" $\Lambda = \beta H^2 +\frac{\alpha}{R^2}$ is applicable  to the total energy density of the universe, 
discuss some cosmological consequences and compare the growing mode of the density contrast with the corresponding mode
of the standard model.
\end{abstract}
\section{Introduction}
The cosmological scale factor obtained integrating the Einstein field equations
, considering the universe as homogeneous and isotropic, has been subject of
study in a numerous quantity 
of works.  Generally, textbooks given more attention to the models with null
pressure 
and absence of a cosmological constant or term.

Sir A. Eddington looked for meaning of the $\Lambda$ term in the Einstein field
equations.  
He declare that $\Lambda $ term corresponds to the absolute energy in a standard
zero condition.
Eddington seems to believe in the necessity of $\Lambda$ as zero point, although
he do not explain
why not consider $\Lambda =0 $, of fact.
Eddington wrote later:  `` $\Lambda $ can not be zero because the zero condition
must correspond  to a possible
rearrangement of the matter of the universe'' \cite{North}.
Otherwise, George Lemaitre considered that the conventional level used to count energy,
taking into account
$\Lambda = 0$, can not be set as fundamental.

In spite of the arguments to preserve the cosmological term, de Sitter and
Einstein joint to a group
that has a opinion in favor of its rejection, in spite of Einstein himself was
one of the pioneer of the 
inclusion of $\Lambda$ in the relativistic  field equations of gravitation.
Moreover, in light of the experimental evidence of the expansion of the
universe, Einstein considered the 
inclusion of $\Lambda $  as `` the biggest blunder of my life'' \cite{Gamov}.

In the present time, the cosmological science indicates that we must have a
non-zero cosmological
term.  Krauss and Turner \cite{Turner} in your work look the return of  the cosmological
constant.
They cited the age of the universe, the formation of the large scale structure
and the matter content of the 
universe as the data that cry out by a cosmological constant.  In another hand,
the data
obtained from the observations of the supernova of type IA, by two independent
teams \cite{Perlmutter}, \cite{Riess},
indicates that the universe has an accelerated expansion.

New problems appear with the inclusion of $\Lambda$ as a constant.  Some of them
are old
today, as the problem about the discrepancy (50 - 120 orders of magnitude) among the observed value for the
present energy density of
the vacuum and the large value suggested by the particle physics standard model 
\cite{Weinberg}.  
An alternative to avoid
this problem is assume that the effective 
cosmological term evolves  and decreases to its present value. 
So, when we refer to $\Lambda$ as 
a cosmological term, we consider that $\Lambda $ is a time-dependent function.

The inclusion of $\Lambda $ term in the field equations is one of the
possibilities
to generate a negative pressure content in the universe, responsible by the
cosmic acceleration.

W. Chen and Y. Wu \cite{Wu} study a cosmological model with a cosmological term
given by $\Lambda \propto R^{-2}$
using a dimensional argument based in conformity with quantum cosmology.  The
authors claim that the `` ansatz''  adopted 
for $\Lambda $ is not in conflict with the observational data and can alleviate
some problems in relation to the inflationary scenario.
Some years after, Carvalho and Lima extended the work of Chen and Wu, including 
$\Lambda = \beta H^2 + \alpha R^{-2} $.
The authors look that the Chen and Wu model do not comprise a negative value for
deceleration parameter ( $q$ ) today,
 but alleviates the density parameter problem.  Otherwise, Carvalho and Lima
\cite{Carvalho} model furnishes $q<0$
and the agreement with nucleosynthesis predictions can be put in terms of the
phenomenological parameters
of the model.  Relaxing the hypothesis that the phenomenological constants of
the model
have the same values during the radiation and matter eras, it is easy to satisfy
the nucleosynthesis constraints.

Posteriorly, John and Joseph \cite{Moncy} argue that the Chen-Wu ``ansatz'' can
be applicable to the total
energy density of the universe, in place of the vacuum density alone.  Since
that  the Planck era is 
characterized by the Planck density , the Chen-Wu `` ansatz'' is better adequate
when applied
 to the total energy density.  The  resulting  model, according the authors, is
free of the horizon, flatness, monopole,
 generation of density perturbations problems.

Our intent in this work is extended the Jonh and Joseph work in a similar sense
that Carvalho and Lima extended
the Chen and Wu work.  We also discuss the growing modes for density contrast and compare 
its evolution with the growing mode of the standard model.
\section{The Model}
Let us consider the universe as homogeneous and isotropic, described by the line
element
\begin{equation}
ds^2 = dt^ 2 - R(t)^2 [\frac{dr^2}{1-\kappa r^2} + r^2(d\theta ^2 +
sin^2(\theta)d \phi ^2)] \, ,
\end{equation}
where $\kappa $ is the curvature parameter.

Describing the universe as a comoving perfect fluid with a $\Lambda $ term, the
Einstein field equations
assumes the form
\begin{eqnarray}
3H^2 + \frac{3\kappa}{R^2} &=& 8\pi G (\rho_m +\rho_{\Lambda})\, ,\\
2\frac{\ddot{R}}{R}+\frac{\dot{R}^2}{R^2}+ \frac{\kappa}{R^2} &=& -8\pi G(P-
\rho_{\Lambda}) \, ,
\end{eqnarray}
where $\rho_m$, $\rho_{\Lambda}$ and $P$ are, respectively, the matter energy
density, the $\Lambda$ energy density
and the pressure.

Extended the Chen-Wu argument to the total density of the universe, we
can write 
\begin{equation}
\rho_m +\rho_{\Lambda}/3 = \xi \, ,
\end{equation}
where $\xi $ is a time dependent function and the $\frac{1}{3}$ in (4) is included for mathematical convenience.
With help of equations (2) (3) and (4) we can write the field equation
\begin{equation}
4\frac{\ddot{R}}{R}-(7+3\nu )(H^2 + \frac{\kappa }{R^2}) + 24 \pi G (1+\nu )\xi
=0 \, .
\end{equation}
The constant $\nu $ in equation (5) is defined by the state equation $P =
\nu \rho $.

Taking into account the ``ansatz'', used by Carvalho et. al \cite{Carvalho} to
the total
energy density of the universe the function $\xi$ assumes the form
\begin{equation}
\xi = \frac{1}{8\pi G}\{\beta H^2 +\frac{\alpha}{R^2} \}\, .
\end{equation}
Therefore, eq. (5) can be written as
\begin{equation}
R\ddot{R}+\alpha _1 \dot{R}^2 + K_ 1 = 0 \, ,
\end{equation}
where $\alpha _1 = -1 +\frac{3(1+\nu)(\beta -1 )}{4}$ and $K _1 = -\kappa
+\frac{3(1+\nu)(\alpha -\kappa )}{4}$.

Integrating equation (7), the solution can be put in the form
\begin{equation}
\pm R^{\alpha _1 +1}\frac{_2 F _1 \{[\frac{1}{2},\frac{\alpha _1 +1}
{2\alpha _1}],[\frac{3\alpha _1 +1}{2\alpha _1}],\frac{R^{2\alpha _1}
K_1}{a}\}}{2(a\alpha _1)^{1/2}[\frac{\alpha _1 +1}{2 \alpha _1}]} = t+b \, ,
\end{equation}
where $_2 F_1$ denotes a hypergeometric function and the integration constants
are denoted by
$a$ and $b$.

The contents of the universe can be considered in the field equations
using the density parameters
\begin{eqnarray}
\Omega _{\Lambda} &\equiv& \frac{\rho _\Lambda}{\rho _c} \nonumber \\
\Omega _{m} &\equiv& \frac{\rho _m}{\rho _c}\nonumber \, ,
\end{eqnarray}
for $\Lambda $ and matter contents, respectively.
The $\rho _c$ is the critical density and is given by $\rho
_c=\frac{3H^2}{8\pi G}$.

Considering the universe flat and using the field equations (2) and (3) we find
\begin{eqnarray}
\Omega _\Lambda &=& \frac{1}{2}\{3-\beta-\frac{\alpha}{\dot{R}^2}\}\\
\Omega _m &=& \frac{1}{2}\{-1+\beta + \frac{\alpha}{\dot{R}^2}\} \, .
\end{eqnarray} 
We can write eqs. (9) and (10) in the form
\begin{equation}
\dot{R}^2 = \frac{2+(1+ \gamma)(1-\beta)}{\alpha(1+\gamma)} \, ,
\end{equation}
where
\begin{equation}
\gamma = \frac{\Omega _\Lambda}{\Omega _m} \, .
\end{equation}
Naturally $\gamma$ depends of the cosmological era.

Rewritten the field equation (7)  as
\begin{equation}
q = \alpha_1 +\frac{K_1}{\dot{R}^2}
\end{equation}
and using expression (11)  we find an expression among the deceleration
parameter ($q$) and
the fenomenological parameters of the model, namely
\begin{equation}
q = -\frac{7-3\beta}{4} +\frac{3\alpha ^2}{4[1-\beta +\frac{2}{1+\gamma}]} \, .
\end{equation}
Although, the expression (14) is interesting from the observational point of view and we obtain it 
without an explicit expression for the scale factor, an explicit form for $R(t)$ is useful  for study the age
and creation of small inhomogeneities in the universe.

With a adequate choose of the parameters we can write the expression (8) for the
scale factor
in a more suitable form.  We guide our choose using equation (13).
Note that assuming $K_1 =0 \longrightarrow q=\alpha _1$.
So, taking into account $K_1$ null we obtain a potential solution for the scale
factor.
\begin{equation}
R(t)= [(1+\alpha _1)t]^{\frac{1}{1+\alpha _1}}\, .
\end{equation}
Consequently the Hubble function and the deceleration parameter are
given, respectively by
\begin{equation}
H=\frac{4}{3(1+\nu)(\beta -1)t}
\end{equation}
and 
\begin{equation}
q=-1+\frac{3(1+\nu)(\beta -1)}{4} .
\end{equation}
Note that, the universe has an accelerated expansion if $1< \beta < \frac{7}{3} $.
However, for $\nu = 0$ the Hubble function $H = \frac{4}{3(\beta -1)t}$ results in a universe older 
than the established by the standard model if $ \beta < 3 $.
So, taking into account  the accelerated expansion of the universe, in accord with the observations from
supernova IA, and the age of the oldest objects in our galaxy that estimates an inferior limit for the age of the universe,
in contradiction with the predictions of the standard model \cite{Chaboyer}
,we can infer a validity range for $\beta $, $1< \beta < \frac{7}{3}$.

For  $K_1 =0 \longrightarrow \alpha = \frac{7\kappa}{3}$, so
 for  different values for the curvature parameter $\kappa $ 
the evolution of the  scale factor (15) is not sensible to the $\kappa $ value.  In other hand, for different
 values for $\beta$ the scale factor evolve differently as show in the
 Fig.1.
\begin{figure}[!ht]
{\includegraphics[width=6cm]{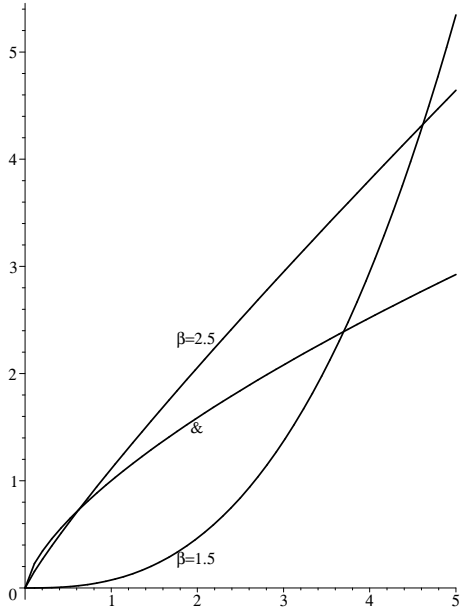}}
\caption{Evolution of the scale factor (15) for $\beta=1.5$ and $\beta=2.5$ and  the scale factor of standard model, signed with $\&$.}
\label{fig:Fig. 1}
\end{figure}

We obtain a coasting type solution \cite{Kolb} if $\alpha_1 = K_1 = 0 $ and for $ \beta =3 $ we obtain a scale factor that mimics
the standard model, although the barionic density differs in the two models.

In other hand, a type de Sitter solution can be obtained substituting $\beta = 1$ in (7), obtaining
\begin{eqnarray}
R(t)=  \frac{a}{2}\{\frac{1}{\exp{\frac{t+b}{\pm a}}} - K1\exp{\frac{t+b}{\pm
a}}\}\, ,
\end{eqnarray}
where the profile is given in the Fig.2
 %
%
%
%
%
\begin{figure}[!ht]
{\includegraphics[width=6cm]{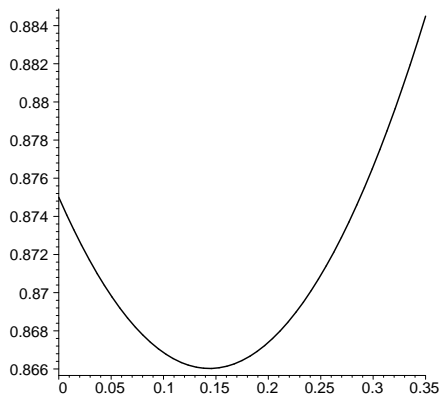}}
\caption{Evolution of the scale factor (18) for $\beta = 1$ and $\alpha = -1$.}
\label{fig:Fig. 2}
\end{figure}
\section{Scalar perturbations}
Although the universe is considered  homogeneous at
large scales, we have inhomogeneities at small scales.  These inhomogeneities are originated from small scalar perturbations
in the homogeneous background. 

The Newtonian formalism is not adequate to study large scale perturbations without invoke procedures which are mathematically
dubious \cite{Pad}.  Although, in regions much smaller than the characteristic length scale, the Newtonian cosmology can be used as an approximation to general relativity and  in such regions there exists a natural choice of coordinates in which Newtonian gravity is applicable.
So, we study the formation of the inhomogeneities in the universe using scalar perturbations in the relativistic framework.  

The relativistic equation,in a universe without $\Lambda $,  for the density contrast  is given by \cite{Pad}:
\begin{equation}
\ddot{\delta}(t)+[2-3(2\nu -v^2)]H\dot{\delta}(t)-\frac{3}{2}H^2\delta(t)(1-6v^2-3\nu^2+8 \nu)=-(\frac{kv}{a})^2 \, ,
\end{equation}
where $\delta $ is the density contrast defined by $\delta = \frac{\delta \rho}{\rho}$ and $\delta \rho $ is the density perturbation.
Considering $\nu = 0$ and $v \sim 0$ and using the field equation for a universe without  a cosmological term $H^2 = \frac{8\pi G \rho}{3}$ we obtain
\begin{equation}
\ddot{\delta}(t) +2H\dot{\delta}(t)-4\pi G\rho \delta(t) =0
\end{equation}
 So, under the circumstances described above the relativistic equation for $\delta $ decay in the  analogue Newtonian equation.

We can obtain the equation for the density contrast in a universe with a cosmological term in a non canonical, but correct  method.
Substituting equation (2) in the equation (20) we obtain
\begin{equation}
\ddot{\delta}(t) +2H\dot{\delta}(t)-\frac{3}{2}H^2\delta(t) +\frac{1}{2}\Lambda \delta(t)=0
\end{equation}
The integration of equation for the density contrast using the scale factor (21) furnishes the modes
\begin{equation}
\delta (t)_{\pm } =t^{\frac{-11+3\beta \pm \sqrt{73-18 \beta +9 \beta ^2}}{6 \beta -6}}\,.
\end{equation}
To have an idea about the evolution of the density contrast in this model compared with the standard model
we define the quantity
\begin{equation}  
f:={\delta_c (t)}-{\delta_s(t_d)}\, ,
\end{equation}
where the sub-script $s$ and $c$ refers to the standard and Chen-Wu models, respectively.
The profile of $f$ is given in the Figure 3.
\begin{figure}[!ht]
{\includegraphics[width=6cm]{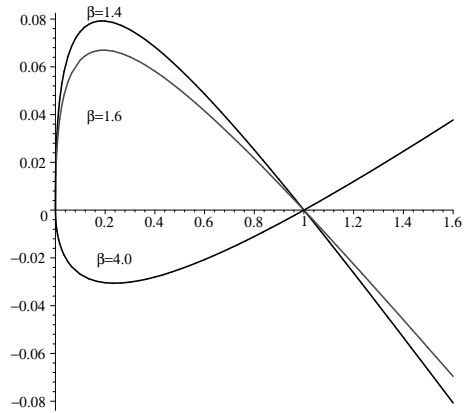}}
\caption{ $f \times t$ for several values of $\beta$.}
\label{fig:Fig. 3}
\end{figure}

Note that at early times the mode $\delta_{c+}$ grows faster than the respective mode for the standard model.
So, in our model we can say that the inhomogeneities are formed more easily and the non linear regime is reached before than
the standard model.

In other hand, using the scale factor (18) and integrating the equation for the density contrast the modes obtained are:
\begin{eqnarray}
\delta_a (t)&\propto& {_2 F_1}\{[\frac{1}{2}+a,\frac{3}{2}+a],[1+2a],-K_1e^{2t}+1\}e^{2t}(-1+K_1e^{2t})^{a-\frac{1}{2}}\, , \\
\delta_b (t)&\propto& {_2 F_1}\{[\frac{3}{2}-a,\frac{1}{2}-a],[1-2a],-K_1e^{2t}+1\}e^{2t}\\
& &(1-K_1e^{2t})^{-2a}(K_1e^{2t}-1)^{a-\frac{1}{2}}\, , \nonumber
\end{eqnarray}
where  $a=\frac{\sqrt{(K_1 -3)K_1}}{2K_1}$.
The profile of the density contrast (24) for $K_1 = -\frac{3}{8}$ is given in the Fig.(4).
\begin{figure}[!ht]
{\includegraphics[width=6cm]{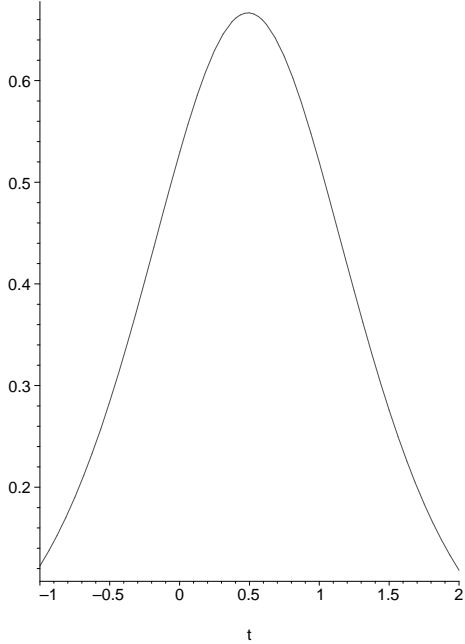}}
\caption{ Evolution for (24) with $K_1 = -\frac{3}{8}$.}
\label{fig:Fig. 4}
\end{figure}
For early times the density contrast (24) grows faster than the standard model.
Note that the density contrast (24) grows with  a posterior decaying, corresponding to the contracting and expansion of the universe governed by the scale factor (18).
For early times the density contrast (24) grows faster than the standard model, as can conclude look
the profile for $f$ in Fig.5.
\begin{figure}[!ht]
{\includegraphics[width=6cm]{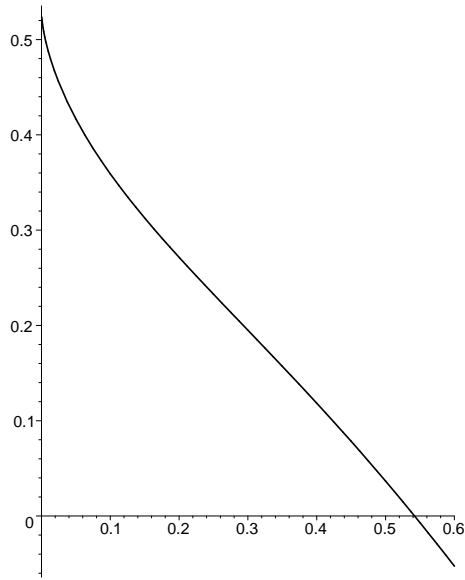}}
\caption{ $f \times t$ }
\label{fig:Fig. 5}
\end{figure}
\section{Conclusions}
In this work we generalized the work of Carvalho and Lima in the same sense that Jonh and Joseph generalized the work of Chen and Wu, previously cited.  Our model do not create conflicts with the observational evidence with respect to the acceleration and age of the universe
and the fenomenological constants of the model can be adjust to the nucleosynthesis predictions. 

Our model of the universe is accelerated and we find the analytical expressions for the growing modes for the density contrast and compare with evolution of the growing mode for  the standard model.

\end{document}